\documentstyle[11pt,epsfig]{article}
\title{
Meson Structure Functions in Valon Model }
\author{ Firooz Arash$^{1,2}$\thanks{email: farash@cic.aut.ac.ir}
\\ $^1${\small Department of Physics, AmirKabir University, Tafresh Campus, Tehran, Iran 15914}\\
$^2${\small Center for Theoretical Physics and Mathematics, AEOI
P.O.Box 11365-8486, Tehran, Iran}\\
}
\begin{document}

\date{\today}
\maketitle
\begin{abstract}
Parton distributions in a {\it{valon}} in the next-to-leading
order is used to determine the patron distributions in pion and
kaon. The validity of the valon model is tested and shown that the
partonic content of the valon is universal and independent of the
valon type. We have evaluated the valon distribution in pion and
kaon, and in particular it is shown that the results are in good
agreement with the experimental data
on pion structure in a wide range of $x=[10^{-4},1]$. 
\end{abstract}
\section{INTRODUCTION}
Unlike the structure function of proton, there are relatively
fewer information on the structure of meson and in particular
pion. The parton distributions in proton have been studied
extensively, both theoretically and experimentally, in a wide
range of $Q^{2}=[0.45, 10000]$ $GeV^{2}$ and $x=[10^{-5}, 1]$.
Yet, pion plays an important role in QCD and its presence is felt
everywhere in hadron physics: pion cloud of nucleon, baryon-meson
fluctuation, and decay of quark to pion-quark which can explain
certain aspects of flavor symmetry breaking in the nucleon sea,
are just a few to name . As a result it is important to determine
and understand its internal structure, which also renders useful
information about nonperturbative QCD. However, at present, the
parton distribution functions of pion are far from being satisfactory.\\
The meson structure is measured in a number of Drell-Yan processes
\cite{1} \cite{2}. Such measurements are concentrated in the
intermediate and large $x$ region, mostly above $0.2$, and hence,
pertinent to the valence quark distribution. Recently, ZEUS
Collaboration have published pion structure function data at  very
low $x$ values from the leading neutron production in $e^{+}p$
collision \cite{3} which provides some details about the sea quark
distribution in pion. Unfortunately, there exists ambiguity in the
normalization of ZEUS data. ZEUS collaboration have used two
different methods to normalize the data. The results differ by a
factor of two. In a recent paper \cite{4} attempts are made to
clarify the normalization ambiguity in the ZEUS data and
independently calculate the pion structure function based on the
valon model. The valon model essentially treats the hadron as the
bound state of its valons. Each valon has its own partonic
structure calculated in QCD. Measurements of Natchmann moments of
proton structure function at Jefferson Laboratory \cite{5} makes
the valon model more credible. The findings of Ref.[5] points to a
new type of scaling which can be interpreted as a constituent form
factor, consistent with the elastic nucleon data. This, in turn,
suggests that the proton structure originates form the elastic
coupling with the extended objects inside the proton \cite{6}. If
confirmed,
such an extended object can be identified as valon. \\
Of course, the notion of structureful objects in hadrons is not
new. Altarelli and Cabibo \cite{7} have used the concept in the
context of $SU(3)\times O(3)$ and R.C. Hwa has termed them as
valons and further developed the concept and showed its
application to many physical processes \cite{8}. It is now well
established that one can perturbatively dress a QCD Lagrangian
field to all orders and construct a structureful object (valon) in
conformity with the color confinement \cite{9} \cite{10}. More
recently, the partonic content of a valon is calculated in the
Next-to-Leading Order (NLO) \cite{11} and shown that if convoluted
with the valon distribution in proton, it gives a fairly accurate
description of the proton structure function data in the entire
kinematical range of measured
values. \\
The underlying assumption of the valon model is that the structure
of a valon is independent of valon type and the hosting hadron.
Therefore, it should provide insight into the structure of hadrons
other than proton, for which the experimental information are
either a rarity or less accurate. In fact, in Ref. [4]  the method
described in Ref. [11] is used to calculate the pion structure
function. However, the focus was on the low $x$ region and the
knowledge gained was pertinent to the sea quark distribution in
the pion; and to provide a resolution of the normalization
ambiguity. Thus, the aim of this paper is three-fold: (a) to
extend the analysis to include the intermediate and large $x$
region; (b) To test the validity of the valon model and extract
the valon distribution of mesons, and (c) to elaborate on the kaon
structure.
\section{The Valon Model}
In the valon model, we assume that baryons and mesons consist of
three, and two valons, respectively. Each valon contains a valence
quark of the same flavor as the valon itself and a sea of partons
(quarks, antiquarks, and gluons). At low enough $Q^{2}$ the
structure of a valon cannot be resolved and the hadron is viewed
as the bound state of its valons. At high $Q^{2}$ the structure of
a valon is described in terms of its partonic content. For a
U-type valon, say, we may write its structure function as
\begin{equation}
F_{2}^{U}(z,Q^2)=\frac{4}{9}z(G_{\frac{u}{U}}+G_{\frac{\bar{u}}{U}})+ \frac{1}
{9}z(G_{\frac{d}{U}}+G_{\frac{\bar{d}}{U}}+G_{\frac{s}{U}}+G_{\frac{\bar{s}}{U}})+...
\end{equation}
where all the functions on the right-hand side are the probability
functions for quarks having momentum fraction $z$ of a U-type
valon at $Q^{2}$. A similar expression can be written for other
types. In Ref. [11] the probability functions, or parton
distributions in the valon are calculated in QCD  to the  NLO at
the scales of $Q_{0}^{2}=0.283$ $GeV^{2}$ and $\Lambda=0.22$ GeV.
Without going into the details, it suffices here to give the
functional forms of them as follows,
\begin{equation}
zq^{valon}_{valence}(z,Q^{2})=a z^{b}(1-z)^{c},
\end{equation}
\begin{equation}
zq^{valon}_{sea}(z,Q^{2}) = \alpha z^{\beta}(1-z)^{\gamma}[1+\eta
z +\xi z^{0.5}].
\end{equation}
The parameters $a$, $b$, $c$, $\alpha$, $\beta$, $\gamma$, $\eta$,
and $\xi$ are functions of $Q^{2}$ and are given in the appendix
of Ref. [11]. Gluon distribution in a valon has an identical form
as in Eq. (3) but with different parameter values. Eqs. (1)-(3)
completely determine the partonic structure of the valon without
any new parameter. We note that the sum rule reflecting the fact
that each valon contains only one valence quark is satisfied for
all $Q^{2}$:
\begin{equation}
\int^{1}_{0} q_{valence}^{vaon}(z,Q^{2})dz=1.
\end{equation}
\section{The Meson Structure Functions}
\subsection{A. Pion}
The determination of parton content of hadron requires the
knowledge of the valon distribution in that hadron. Let us denote
the probability of finding a valon carrying momentum fraction $y$
of the hadron by $G_{\frac{valon}{h}}(y)$, which describes the
wave function of hadron in the valon representation, containing
all the complications due to confinement. Following [4, 8, 11] and
\cite{12}, we write the valon distribution in a meson as:
\begin{equation}
G_{\frac{valon}{meson}} (y)=\frac{1}{\beta[\mu_{m}+1,\nu_{m}+1]}y^{\mu_{m}}(1-y)^{\nu_{m}}.
\end{equation}
with the requirements that the above form satisfies the number and momentum
sum rules:
\begin{equation}
\int^{1}_{0} G_{\frac{valon}{meson}}dy=1
\hspace{3cm}\sum_{valon}\int^{1}_{0} y G_{\frac{valon}{meson}}dy=1
\end{equation}
where, $\beta[i,j]$ is the Euler beta function and
$G_{\frac{valon}{h}}(y)$ stands for the distribution of a U-valon
in $\pi^{+}$ or a D-valon in $\pi^{-}$. By interchange of
$\mu\leftrightarrow \nu$ the anti-valon distribution in the same
meson is obtained. \\
An essential property of the valon model is that the structure of
hadron in the valon representation is independent of the probe.
This means that the parton distribution in a hadron can be written
as the convolution of the partons in the valon and the valon
distribution in the hadron. For the case of pion, this translates
into:
\begin{equation}
xu^{\pi^{+}}_{valence}(x,Q^{2})=\int^{1}_{x}dy
\frac{x}{y}G_{\frac{U}{\pi^{+}}}(y)u_{\frac{valence}{U}}({\frac{x}{y}},Q^{2})
\end{equation}
\begin{equation}
x\bar{d}^{\pi^{+}}_{valence}(x,Q^{2})=\int^{1}_{x}dy
\frac{x}{y}G_{\frac{\bar{D}}{\pi^{+}}}(y){\bar{d}}_{\frac{valence}{\bar{D}}}({\frac{x}{y}},Q^{2})
\end{equation}
\begin{equation}
xd^{\pi^{-}}_{valence}(x,Q^{2})=\int^{1}_{x}dy
\frac{x}{y}G_{\frac{D}{\pi^{-}}}(y)d_{\frac{valence}{D}}({\frac{x}{y}},Q^{2})
\end{equation}
\begin{equation}
x\bar{u}^{\pi^{-}}_{valence}(x,Q^{2})=\int^{1}_{x}dy
\frac{x}{y}G_{\frac{\bar{U}}{\pi^{-}}}(y){\bar{u}}_{\frac{valence}{\bar{U}}}({\frac{x}{y}},Q^{2})
\end{equation}
As for the sea quark distribution , we will take the example of
$\pi^{+}$. $\pi^{+}$ has two valons, $U$ and $\bar{D}$ and each
contributes to the sea and gluon content of pion as:
\begin{equation}
xq^{\pi^{+}}_{sea}(x,Q^{2})=\int^{1}_{x}dy\frac{x}{y}
G_{\frac{U}{\pi^{+}}}(y)q_{\frac{sea}{U}}(\frac{x}{y},Q^{2})+
\int^{1}_{x}dy\frac{x}{y}
G_{\frac{\bar{D}}{\pi^{+}}}(y)q_{\frac{sea}{\bar{D}}}(\frac{x}{y},Q^{2})
\end{equation}
Evaluation of these convoluted integrals requires us to determine
$G_{\frac{valon}{\pi}}(y)$ or, alternatively, finding $\mu_{m}$
and $\nu_{m}$. Since the two valons in the pion, apart from
flavor, cannot be distinguished and since the masses of the $U$
and $\bar{D}$ valons are the same, therefore their average
momentum also must be the same. This can be achieved only if
$\mu_{m}=\nu_{m}$, leaving us with only one parameter. We use the
valence distribution data in $\pi^{-}$ at $Q^{2}=25$ $GeV^{2}$
from the Drell-Yan experiment of E615 collaboration [2] to find
this parameter. The fit to the data of Ref. [2] is shown in Fig.
(1), which is obtained by taking the starting
scale,$Q^{2}_{0}=0.47$ $GeV^{2}$ with $\Lambda_{QCD}=0.22$ GeV.
\begin{figure}
\epsfig{figure=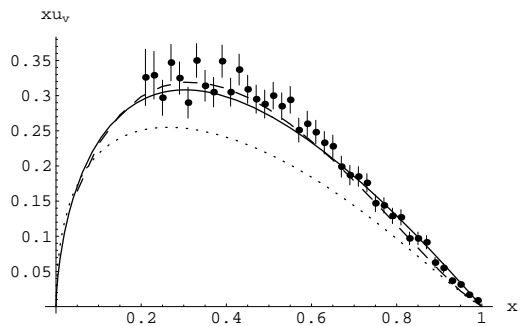,width=7cm}
\caption{\footnotesize Comparison of the pion valence
distribution,$u_{valence}^{\pi^{-}}(x)$, result from the valon
model calculation (solid line), SMRS, Ref. [13] (dashed line),
GRS, Ref. [14] (dotted line), and the data from E615 [2] at
$Q^{2}=25 GeV^{2}$.} \label{fig1}
\end{figure}

The goodness of the fit is checked by $\chi^{2}$ minimization
procedure. We find that the $\chi^{2}$ per degree of freedom is
$1.2$ and that $\mu_{m}=\nu_{m}=0.01$. As a comparison, in Fig.(1)
we have also shown the results from the determination of \cite{13}
as dashed line and those of \cite{14} as dotted line, both at
$Q^{2}=25$ $GeV^{2}$. In fitting the data we have also allowed
both $\Lambda_{QCD}$ and $Q^{2}_{0}$ to be free parameters and
obtained an equally good description of the data with
$Q^{2}_{0}=0.35$ $GeV^{2}$ and $\Lambda_{QCD}=0.175$ GeV resulting
in $\mu_{m}=\nu_{m}=0.03$. However, it seems that
$\Lambda_{QCD}=0.175$ is too low for four active flavor. As such,
we choose the former against the latter. With the Determination of
these parameters, it is important to check the valence quark sum
rule in pion. Using the explicit form of Eq. (2), i.e.
$zq_{valence}^{valon}(z=x/y,Q^{2})$, from the appendix of Ref.
[11], we find that each of the integrals in Eqs. (7-10) gives
$0.9994$, $1.002$, and $1.008$ at $Q^{2}=3, 10, 20$ $GeV^{2}$,
respectively; an excellent confirmation of the valence quark sum
rule. The first two moments of pion valence quark distribution is
also calculated at $Q^{2}=49$ $GeV^{2}$ for the purpose of
comparison with the findings of Ref. [13]. We find that
$2<xq_\frac{valence}{\pi}>=0.378$ and
$2<x^2q_\frac{valence}{\pi}>=0.151$. This is to be compared with
$0.40\pm 02$ and $0.16 \pm 0.01$ of Ref. [13], respectively.

From Eq. (5) we can infer some knowledge about the charge and
matter distributions of the pion [8]. The longitudinal momentum
space is related to the coordinate space by a Fourier Transform
and a boost to infinite-momentum frame. Since the valon structure
originates from QCD virtual processes, which are flavor
independent, the matter and the charge densities in the valon
ought to be flavor independent. Hence, the charge density in, say,
$\pi^{+}$ is
\begin{equation}
\rho_{q}(y)=\frac{2}{3}G_{\frac{U}{\pi^{+}}}(y)
+\frac{1}{3}G_{\frac{\bar{D}}{\pi^{+}}}(y),
\end{equation}
whereas for the matter density distribution we assume that it is proportional to
the total valon distribution, i.e. ,
\begin{equation}
\rho_{m}(y)=\frac{1}{2}(G_{\frac{U}{\pi^{+}}}(y)+G_{\frac{\bar{D}}{\pi^{+}}}(y)).
\end{equation}
The integral of both quantities in Eqs.(12) and (13) are equal to
one. \\
Let us write the explicit form of $G_{\frac{valon}{\pi^{+}}}$:
\begin{equation}
G_{\frac{U}{\pi^{+}}}=G_{\frac{\bar{D}}{\pi^{+}}}=1.020(1-y)^{0.01}y^{0.01}.
\end{equation}
It is evident that this function is very broad in momentum space.
This feature is expected, for, it indicates that the valons are
tightly bound. This is also a reflection that the pion is much
lighter than the mass of its constituent quarks. The parameters,
$\mu_{m}$ and $\nu_{m}$ obtained here are slightly different from
those used in Ref. [4,11,12] and are significantly different from
those quoted in \cite{15}. In [4,11] the values $\mu_{m}=0.01$ and
$\nu_{m}=0.06$ were used, while the determination of Ref. [12] is
$\mu_{m}=\nu_{m}=0$, thus, there is no significant differences
among this work and those of Ref.[4,11,12].  In Ref. [15] the pion
cloud model in conjunction with the valon model is used to
calculate the pion structure. They find that $\mu_{m}=0.044$ and
$\nu_{m}=0.372$. In this work and in Ref. [4] the parton
distribution in a valon is derived from QCD alone, with no
phenomenological assumption. However, in Ref. [4], the focus was
on the low $x$ behavior of the pion structure function, whereas
here we have used data at rather large $x$ ($x> 0.2$) region to
determine the valon distribution in the pion. In Fig. (2) we
present the full pion structure function at $Q^{2}=7$ and
$Q^{2}=15$ $GeV^{2}$ at small $x$, ($x=[10^{-4}, 10^{-2}]$),
region and compared the results with those obtained from
Sutton,{\it{et al.}} \cite{13} and Gluck, Reya, and Schienbein
\cite{14} parameterizations. In Fig. (2) two sets of data points
are shown, which correspond to the two different methods of
normalization used by ZEUS collaboration.While the results from
Ref. [14] agrees with the additive quark model normalization, the
parameterization of Ref. [13] is qualitatively closer to the
effective one-pion-exchange model normalization. Note that
although the parameterization of Ref.[14] provides a good
description of the additive quark model normalization of the data,
it fails to describe the large $x$ data of Ref. [2] as is apparent
from Fig. (1). As can be seen from the figure, the valon model
results are in good agreement with the pion flux normalization of
the data. It is also interesting to note that in the valon model a
simple relationship holds rather well between proton and pion
structure functions, namely, $F_{2}^{\pi}=k F_{2}^{p}$, with
$k\simeq 0.37$ [4]. This observation has also been made by the
ZEUS Collaboration [3]. The ZEUS group will soon release two new
measurement: one is photoproduction study from which the pion
trajectory can be determined; and the second is a measure of the
exponential $P_{t}$ slopes in deep inelastic scattering that can
be used to limit the choice for the form factor $F(t)$ \cite{16}.
These measurements should help to resolve the normalization issue.
\begin{figure}
\centerline{\begin{tabular}{ccc}
\epsfig{figure=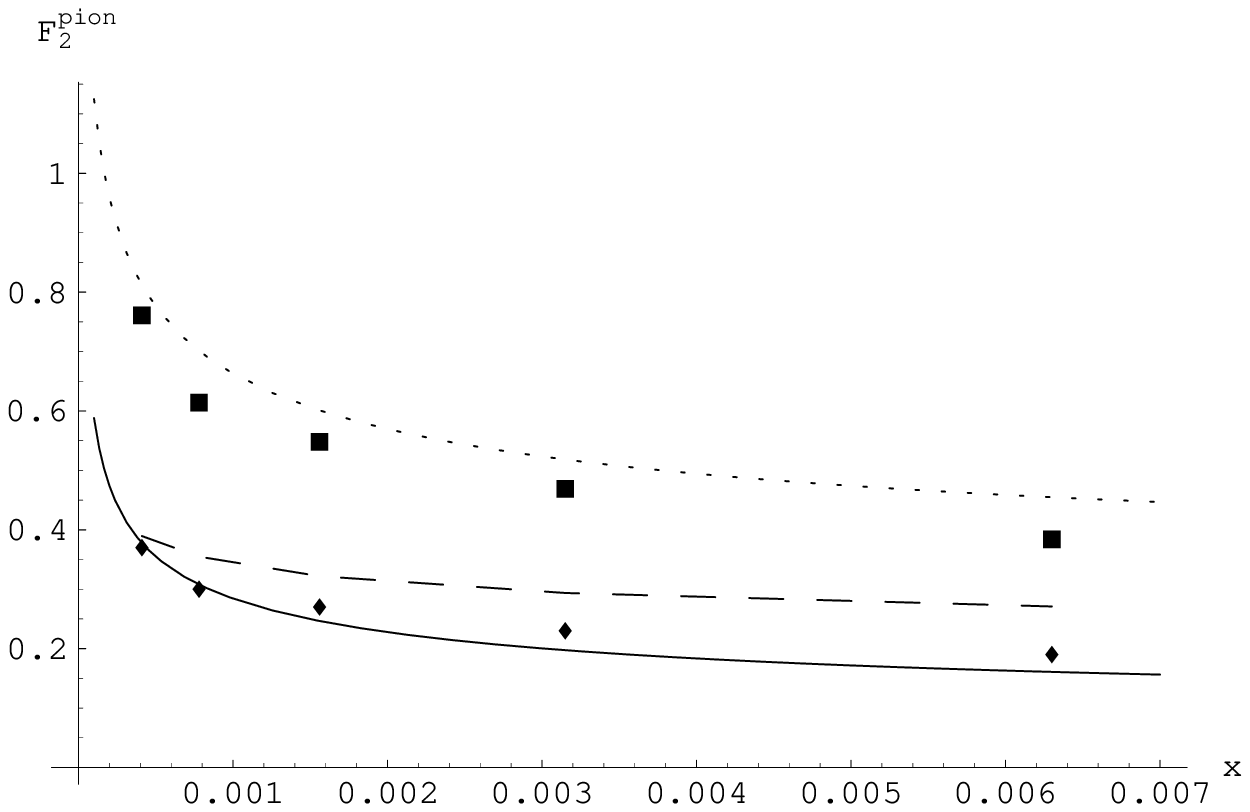,width=7cm}
 &\hspace{2cm}&
\epsfig{figure=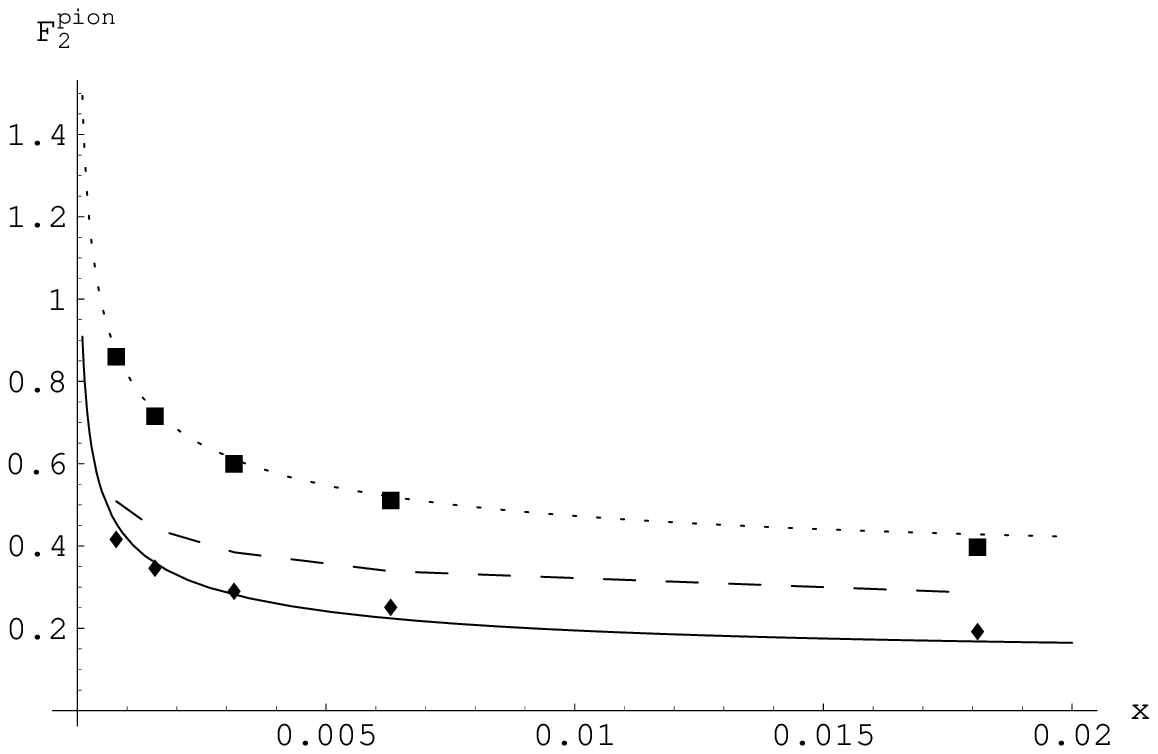,width=7cm}
\end{tabular}  }
\caption{\footnotesize Pion structure function at $Q^{2}=7
GeV^{2}$ (Left) and at $Q^{2}=15 GeV^{2}$ (Right). The diamonds
and squares are pion flux and additive quark model normalization
of the data [3], respectively. The solid line represents the
calculated results from the valon model. The bdashed line is the
result from SMRS [13] and the dotted line corresponds to GRS
[14]determination.} \label{fig2}
\end{figure}
\subsection{B. Kaon}
The treatment of the kaon structure function is similar to that of
pion, except that we need to determine valon distribution in the
kaon. We will concentrate on $K^{-}$, since there are some data
\cite{17} which provide information about the valence distribution
in $K^{-}$. In Refs. [8, 11, 12] it is stated that the general
form of the exclusive valon distribution in a meson is as follows
\begin{equation}
G_{V_{1}V_{2}}(y_{1},y_{2})=[\beta(\mu_{k}+1,\nu_{k}+1)]^{-1}y_{1}^{\mu_{k}}y_{2}^{\nu_{k}}\delta(y_{1}+y_{2}-1)
\end{equation}
Integrating over either of $y_{i}$ will give the individual valon
distribution in the meson, as in Eq. (5). So, we can write a
similar equation for the valon distributions in $K^{-}$, except
that $\mu_{m}$ and $\nu_{m}$ are different and we label them as
$\mu_{k}$ and $\nu_{k}$, respectively. One way to determine these
parameters is fitting $G_{\frac{valon}{hadron}}(y)$ to some
experimental data. Unfortunately, there is no experimental data on
the valon distribution of any hadron, and from the theoretical
point of view, the function $G_{\frac{valon}{hadron}}(y)$ cannot
be evaluated accurately. However, there are some limited data on
the ratio
\begin{equation}
R=\frac{x\bar{u}_{k^{-}}}{x\bar{u}_{\pi^{-}}}
\end{equation}
 at large $x$ values \cite{17}, and can be used to
 fit equations of the form (7-10) to get $\mu_{k}$ and $\nu_{k}$.
 Maintaining the same values for $Q_{0}^{2}$ and $\Lambda$ as in
 the case of pion, we have fit the data of Ref. [17] and obtained $\mu_{k}=0.13$
 and $\nu_{k}=0.28$ with a $\chi^2=0.646$ per degrees of freedom.
 The fit is shown in Fig. (3). In our fit to this data set, both
 valence and sea ${\bar{u}}$ are included, although the data
 points are at rather large $x$ and hence, the sea quark
 contribution to $R$ is marginal.
 \begin{figure}
\epsfig{figure=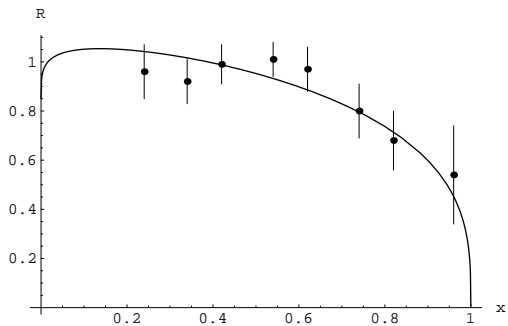,width=7cm}
\caption{\footnotesize The ratio $R$ as a function of $x$ as
measured by [17]. The curve is obtained from the model by fitting
the data at $Q^{2}=25$ $GeV^{2}$.} \label{fig3}
\end{figure}
In the alternative, we can be guided by
 making a simple phenomenological assumption as follows [12]. Since
 $K^{-}$ consists of two valons with different masses. It is
 obvious from Eq. (15) that the average momentum fractions of the
 light valon, $\bar{y_{1}}$ and the heavy valon, $\bar{y_{2}}$, are
 $\bar{y_{1}}=(\mu_{k}+1)/(\mu_{k}+\nu_{k})$ and $\bar{y_{2}}=(\nu_{k}+1)/(\mu_{k}+\nu_{k})$, respectively.
 Thus, letting the ratio of the momenta be equal to ratios of their masses, we
 get
 \begin{equation}
 \frac{\bar{y_{1}}}{\bar{y_{2}}}=\frac{m_{U}}{m_{S}}\simeq
 \frac{300}{500}=\frac{\mu_{k}+1}{\nu_{k}+1}=0.6
 \end{equation}
 In this way $\nu_{k}$ is restricted by
 \begin{equation}
 \nu_{k}=(\mu_{k}+0.4)/0.6,
 \end{equation}
 hence, leaving us with only one parameter, $\mu_{k}$, to be determined.
 A one parameter fit to the data of Ref. [17] is highly
 unsatisfactory, with $\mu_{k}=-0.35$ as can be seen in Fig. (4). Authors of  Ref. [12]
 have used inclusive distribution of $K^{+}\rightarrow \pi^{+}$ and
$K^{+}\rightarrow \pi^{-}$ and found $\mu_{k}=1$ and $\nu_{k}=2$
which also do not support the data of Ref. [17]. An earlier QCD
calculation \cite{18} of the ratio, which takes into account the
difference in quark mass, however agrees with the data.
Nevertheless, it seems that $R$ being around 0.5 at $x=0.95$ is
too high. Considering the large error bars in the data, the
accuracy of the data, at least, at large $x$ is suspect. One
possible way of determining the valence quark density in kaon
would be the difference of two cross sections in $K^{+}
p\rightarrow \mu^{+}\mu^{-}+ X$ and $K^{-} p\rightarrow
\mu^{+}\mu^{-}+ X$ processes.

 \begin{figure}
\epsfig{figure=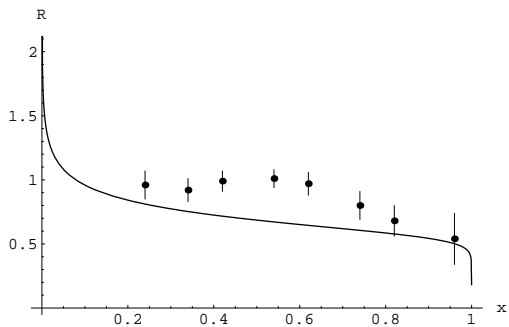,width=7cm}
\caption{\footnotesize The ratio $R$ as a function of $x$ as
measured by [17]. the curve is obtained with $\mu_{k}=-0.35$ at
$Q^{2}=25$ $GeV^{2}$.} \label{fig4}
\end{figure}
While the determination of the valon distribution in kaon remains
uncertain, we will proceed with the values obtained for $\mu_{k}$
and $\mu_{k}$ without the restriction of Eq. (18), namely,
$\mu_{k}=0.13$ and $\nu_{k}=0.28$ and arrive at the following
valon distributions in $K^{-}$;
\begin{equation}
G_{\bar{U}}^{k^{-}}(y)=1.4768y^{0.13} (1-y)^{0.28},
\hspace{1.5cm}G_{S}^{k^{-}}(y)=1.4768y^{0.28} (1-y)^{0.13}
\end{equation}
With these relations the average momentum fraction carried by
$\bar{U}$ and $S$ valons are 0.469 and 0.531, respectively. Using
these valon distributions along with Eqs. (7-10) for $K^{-}$, the
valence quark distributions for $\pi^{-}$ and $k^{-}$ are
calculated at $Q^{2}=25 GeV^{2}$ and shown in Fig.(5).

 \begin{figure}
\epsfig{figure=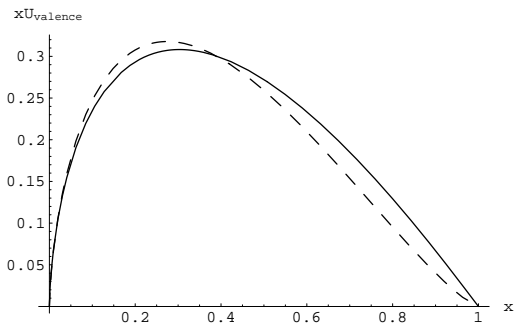,width=7cm}
\caption{\footnotesize The valence quark distribution,
$x\bar{u}(x)$, in $\pi^{-}$ (solid curve) and in $K^{-}$ (dashed
curve) . The curves are obtained from the model at $Q^{2}=25$
$GeV^{2}$.} \label{fig5}
\end{figure}
It should be noted that in Drell-Yan processes, obtaining
experimental information on the strange valence and sea quark
distribution in kaon is not practical. Because those components in
kaon only contribute to the total cross section of the Drell-Yan
processes through valence-sea and $s-\bar{s}$ annihilation and
these contributions being small makes it difficult to separate
them. The valon model, on the other hand, provides valuable
information about these components. Figures (6) shows the strange
valence and sea quark distributions in $K^{-}$. Both are
calculated at $Q^{2}=25$ $GeV^{2}$ from the model. For the purpose
of comparison, in Figure (7) we show sea quark distributions in a
valon. The sea quark distribution is calculated originally for $u$
and $d$ type partons. The strange sea quark distributions are
generally smaller than the light sea quark distributions so that
in proton $2{\bar{s}}/({\bar{u}}+{\bar{d}})\sim 0.5$\cite{19}
Nevertheless, by the time $x\sim 10^{-4}$ ${\bar{s}} \approx
{\bar{u}}$.

\begin{figure}
\centerline{\begin{tabular}{ccc}
\epsfig{figure=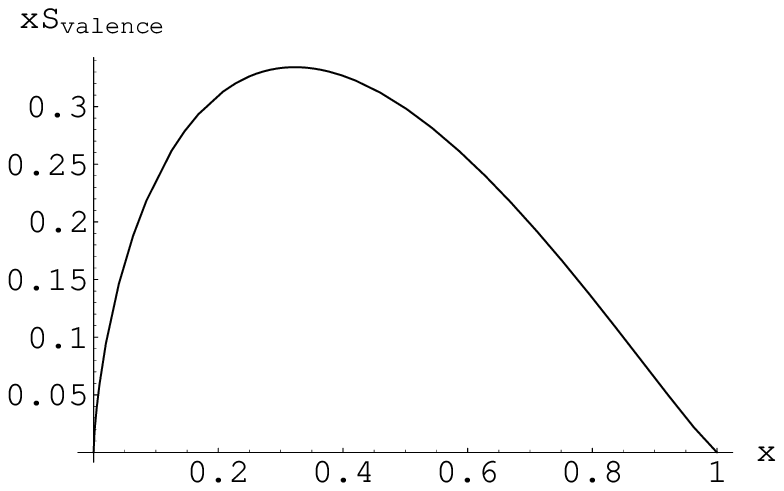,width=7cm}
 &\hspace{2cm}&
\epsfig{figure=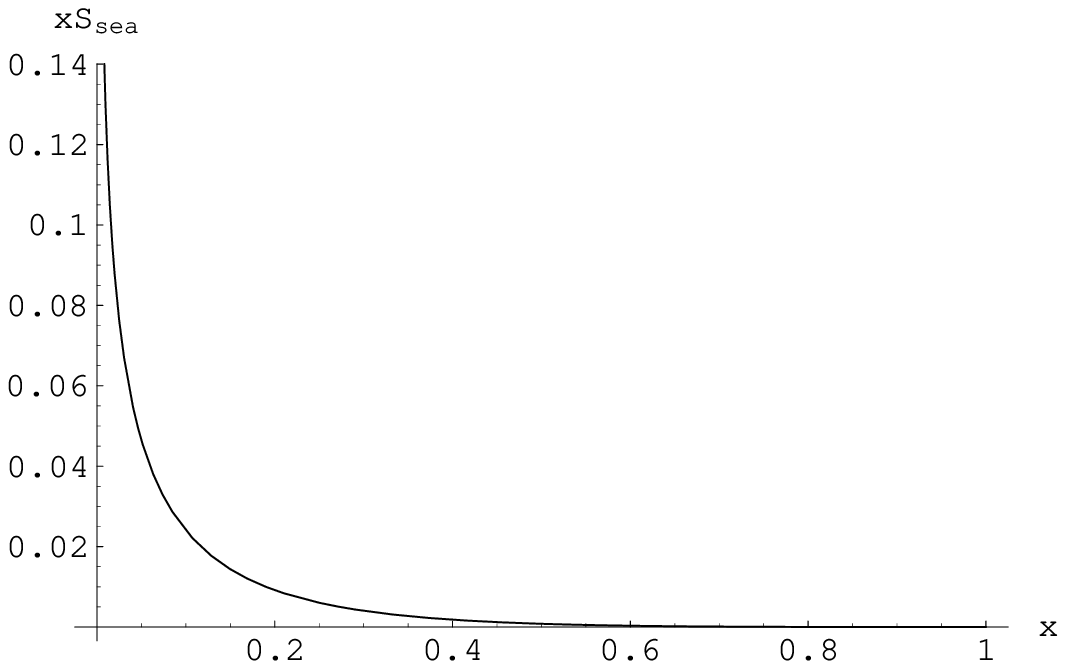,width=7cm}
\end{tabular}}
\caption{\footnotesize Strange quark distribution in $K^{-}$ at .
$Q^{2}=25$ $GeV^{2}$. Left figure is the valence component and the
right figure is that of strange sea distribution. } \label{fig6}
\end{figure}
 \begin{figure}
\epsfig{figure=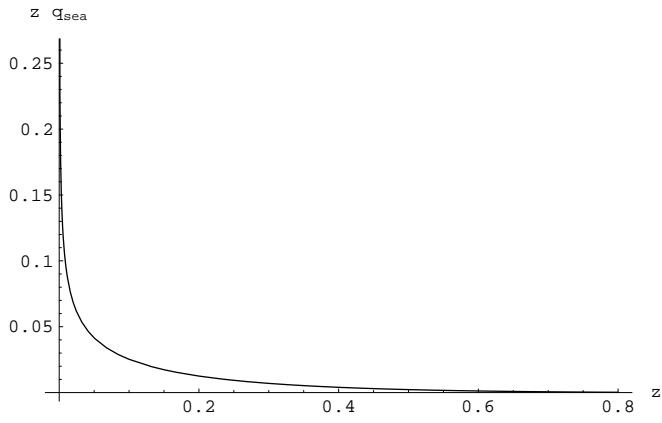,width=9cm}
\caption{\footnotesize Sea parton distributions in a valon at
$Q^{2}=25$ $GeV^{2}$.} \label{fig7}
\end{figure}
\section{Conclusion}
We have used the notion of the valon model to determine the
structure of mesons. Since the partonic structure of a valon is
already known, the structure function of any meson needs only two
parameters to be completely determined. The valon model provides
information about the sea quark distribution of mesons that are
out of reach experimentally. While the pion structure is
determined fairly accurately, the structure of kaon remains
uncertain, due to the lack of accurate data or incompatibility of
data sets from different experiments.

 \end{document}